\RequirePackage{fix-cm}
\documentclass[smallextended]{svjour3}       
\smartqed  
\usepackage{graphicx}
\usepackage[caption=false]{subfig}
\usepackage{nicefrac}
\usepackage{makecell}
\usepackage{amssymb}
\usepackage{amsmath}
\journalname{Experimental Astronomy}
\begin{document}

\title{On the extension of the sensitive area of an extensive air shower surface array}

\titlerunning{Extension of the sensitive area of an EAS array}        

\author{H. Hedayati Kh.}

\institute{H. Hedayati \at
              Department of Physics, K.N. Toosi University of Technology, P.O. Box 15875-4416, Tehran, Iran \\
              Fax: +98-21-23064218\\
              \email{hedayati@kntu.ac.ir}
}

\date{Received: date / Accepted: date}

\maketitle

\begin{abstract}
A large distance between true and reconstructed core locations of an extensive air shower (EAS) may result in great systematic mis-estimation of EAS parameters. The reconstruction of those EASs whose core locations are outside the boundary of a surface array (outside EAS (OEAS)) results in a large distance of the reconstructed core location from the true one, especially when the true core is far outside the array. Although it may not be mentioned, the rejection of OEASs is a necessary and important step in the reconstruction procedure of an EAS. In this paper, an existing technique is optimized for the rejection of OEASs. The simultaneous use of this technique and a recently developed approach for reconstructing the core location of an EAS can significantly increase the sensitive area of a surface array.
\keywords{Cosmic rays \and Extensive air showers}
\PACS{96.50.S- \and 96.50.sd}
\end{abstract}

\section{Introduction}
\label{intro}
In a surface array during an EAS event, a particularly minimum number of array detectors should be usually triggered to record the event (threshold condition). Sometimes, an EAS with the true core outside the array boundary can satisfy the threshold condition. Reconstructed core location of these EASs not only is inside the array, but may also have considerable distance from array boundary, especially for those EASs with the true core far outside the array.\\
This problem may sometimes be even worse. For some OEASs, a few number of array detectors may exists that, in spite of the large distance from the true core location, detect a significant number of particles. Often, the reason of this event is a single particle of the EAS that moves behind the EAS front and, as a result of Landau fluctuation \textbf{ and/or} a cascade in detector material, generates a large pulse height which is mistaken for a high particle density location. Corrupted data of these detectors in addition to destructive effect on the reconstruction of the EAS direction may cause difficulty on the rejection of such EASs as OEASs even with sophisticated core location reconstructing algorithms. Therefore, certain quality cuts should be applied to safely recognize and distinguish good events from badly reconstructed ones.\\
In short, rejection of OEASs is very important, since if the distance of the reconstructed core location from the true core of an EAS is large, in addition to a systematic tilt in reconstructed arrival direction of the shower, other reconstructed parameters such as shower size, age parameter, etc. will have significant systematic errors.\\
There are various methods to reject OEASs in a surface array. A sophisticated method for the rejection of OEASs is using complementary data other than those of the surface array detectors alone (e.g. data of Cherenkov light detectors, \cite{Arqueros:1999uq}). The disadvantages of these techniques are that they raise the cost of the array construction and, in some cases such as air Cherenkov detectors, restrict the duty cycle of the array considerably (e.g. to a dark clear moonless night).\\
If one merely relies on data of a surface array detector, the most common method is the border cut (e.g. KASCADE \cite{antoni2003cosmic}): Based on this approach, the reconstructed core and the first guess core position have to be deep inside the boundary of the array. Furthermore, it is required that the station containing the largest signal is not on the border of the array. Although the marginal area of an array is a narrow region, it usually has a considerable contribution to the total area of the array. So, this technique reduces the \textbf{effective area of the array.} Also, it can reduce the energy extent of detected EASs by the array, because the most energetic EASs which can satisfy the threshold condition are those EASs with the true core near the boundary \textbf{(see figure 6 of \cite{hedayatiA})}.\\
An interesting technique for the rejection of OEASs is to find the weighted mean of distances of the registered particles from the reconstructed shower core \cite{krawczynski1996optimized}, as named by authors as $r_p$:
\begin{equation}
r_p=\frac{\sum\limits^{N}_{i=1}{\alpha_i n_i  r_i}}{\sum\limits^{N}_{i=1}{\alpha_i n_i}}
\label{Rp}
\end{equation}
The summation includes all triggered detectors in the array. The parameter $r_i$ represents the distance of the $i$th detector from the center of gravity (COG) of the responding detectors, $n_i$ the particle density measured by the $i$th detector, and $\alpha_i$ a weight which takes into account the inhomogeneous detector spacing in an asymmetrical surface array. The weights are inversely proportional to the density of detectors around the $i$th detector. According to the authors, exceptionally large distances between the true and reconstructed shower cores result in exceptionally large $r_p$ values, so OEASs can be identified by their $r_p$ values (actually we will see below that when the true core is outside the array, contrary to authors' view, using distances from the true core location as $r_i$s results in the greatest $r_p$s). So, when we use a common method for reconstructing the core location, this technique can only identify far OEASs and still need a relatively large border cut.\\
Sophisticated techniques such as neural networks \cite{mayer1992neural} cannot drastically improve the above methods and the true core location must be inside array and have significant distance from the array border in order to be reconstructed reliably.\\
In this paper, $r_p$ parameter will be optimized for the rejection of OEASs. This rejection technique relies only on the surface array data and can increase sensitive area of an array by maintaining the marginal EASs (internal EASs with a core location near boundaries). In spite of the fact that deterioration in energy resolution and systematic bias of marginal events occur because of core location mis-reconstruction, if their core locations can be reconstructed with an acceptable resolution (as a new method called SIMEFIC II \cite{hedayatiB} suggests, it will be briefly reviewed below), the situation will get better.\\
For the real EASs, we do not have the exact core location, so they are not suitable for comparing the results of different methods for the rejection of OEASs. Therefore, in order to prove the functionality of this technique, simulated EASs whose specifications are introduced in the next section are used.
\begin{table*}[h!]
\centering
\begin{tabular}{| l | c |} 
 \hline
 Specification & Values and Ranges\\ [0.5ex] 
 \hline\hline
 energy range of primaries & 50 $\textrm{TeV}<E<$ 5 PeV\\\
zenith angle range of primaries & $0^\circ<\theta<60^\circ$
\\
azimuth angle range of primaries & $0^\circ<\theta<180^\circ$
\\
 geographical longitude & 51 E  \\ 
 geographical latitude & 35 N  \\
 altitude & 1200 m  \\
 earth magnetic field ($B_x$) & $28.1 \,\mu$T  \\
 earth magnetic field ($B_z$) & $38.4 \,\mu$T \\
 low energy hadronic model & Fluka 2011.2b \cite{ferrari2005fluka}  \\
 high energy hadronic model & QGSJETII-04 \cite{ostapchenko2011monte}  \\ [1ex] 
 \hline
\end{tabular}
\caption{EASs' specifications. The primary particle of 90\% of the showers are protons and the remaining primary particles are alphas. Other specifications are CORSIKA default values.}
\label{CORSIKADef}
\end{table*}
\section{Air Shower Simulations}
\label{sec:SimSec}
In order to confirm the performance of OEAS's identification parameter, more than 400,000 CORSIKA version 7.4 \cite{heck1998corsika} simulated EASs whose specifications are summarized in Table ~\ref{CORSIKADef} were generated.\\
\begin{figure}
\centering
\resizebox{0.7\hsize}{!}
{\includegraphics[width=\hsize]{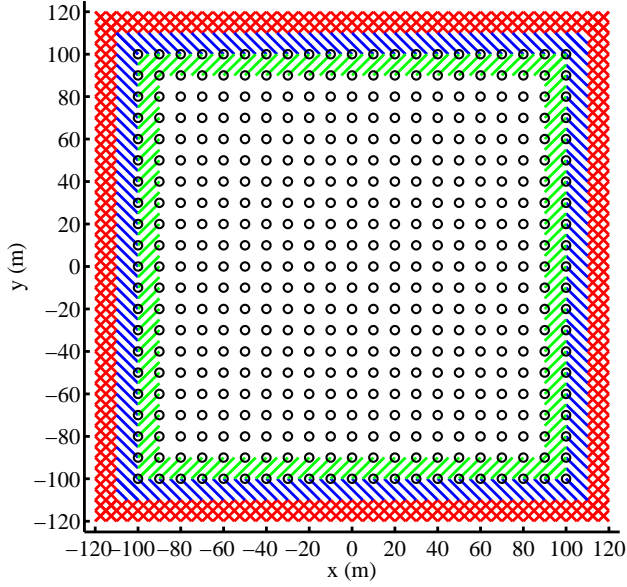}}
\caption{Layout of the assumed array. The positions of detectors in the array are shown by empty black circles (not to scale). The last internal ring of the array is hatched by green lines. Also, the first and second external rings are shown in this figure.}
\label{1_array}
\end{figure}
A hypothetical surface array, similar to that of \cite{hedayatiA} (a square array with $21\times21$ detectors, the network constant of $10$ m over the total area of $200\times200 \textrm{m}^2$) is applied (Fig.~\ref{1_array}). A threshold condition of triggering at least 12\% (53 detectors) of array detectors is exerted. For finding the arrival direction, plane front approximation is used. More details about array and detectors can be found in \cite{hedayatiA}.\\
In this study, the detector response has not been addressed. However, it should be emphasized that, the detector response plays an important role, especially when the detected number of particles is small and the fluctuations in the detector response may be important. Therefore, in a more detailed study, the detector response should be taken into account.
\section{A brief review of the SIMEFIC II method}
In the SIMEFIC II method, the core location of an EAS is reconstructed with the weighted center of gravity (WCG):
\begin{equation}
\begin{aligned}
  x_{WCG}=\frac{\sum_{i=1}^{N}{x_iw_i}}{\sum_{i=1}^{N}{w_i}},  \\
  y_{WCG}=\frac{\sum_{i=1}^{N}{y_iw_i}}{\sum_{i=1}^{N}{w_i}},
\end{aligned}
\end{equation}
where $w_i$’s are weights of fired detectors (FD) of the array during an EAS event, $N$ is the number of FDs, and $x_i$ and $y_i$ are the locations of $i$th FD. The weight of a FD, $w_i$, is defined as $n_in_j/d_{3ij}$, where $n_i$ is the signal heights of the $i$th FD, $n_j$ is the signal height of a $j_th$ detector (a detector near the $i$th detector; to find the details on how to choose a detector near the $i$th detector as the $j$th detector, the reader may consult \cite{hedayatiB}), and $d_{3ij}$ is the 3D distance between $i$th and $j$th FDs. The 3D distance between two detectors is defined as the distance between the first crossing particles from each detector. It can be found by the combination of detector location and its recorded time, as explained in \cite{hedayatiB}.\\
While the above relation for finding the core location is good enough for those EASs that impact in the central region of the surface array, its precision becomes worse for those EASs which impact in the outer region of the array, especially those EASs with true core near boundaries (marginal EASs), because for the marginal EASs, those FDs far from the core which are fired only on one side of the true core, displace the reconstructed core in their average directions (their average direction is approximately in the direction of the center of the array from the true core location). Of course, the same problem arises for every core location reconstruction method based on a kind of weighted COG of FDs (e.g. simple COG).\\
In order to improve this asymmetry, we divide $N$ FDs into two sets: the first $M$ highest weighted FDs and the remaining $N-M$ FDs. Then, we find an initial core location using the $M$ ($1<M<N$) highest weighted FDs by the following relation:
\begin{equation}
\begin{aligned}
x_{rc}^M=\frac{\sum_{i=1}^M{x_iw_i}}{\sum_{i=1}^M{w_i}},\\
y_{rc}^M=\frac{\sum_{i=1}^M{y_iw_i}}{\sum_{i=1}^M{w_i}},
\end{aligned}
\label{WCGM}
\end{equation}
The $N-M$ remaining FDs are used for the correction of this initial core location as:
\begin{equation}
\begin{aligned}
x_{ct}=&\frac{\sum_{i=M+1}^N{(x_{rc}^M-x_i)w_i}}{\sum_{i=1}^N{w_i}}\\
=&\frac{\sum_{i=M+1}^N{(x_{rc}^M-x_i)w_i}}{\sum_{i=M+1}^N{w_i}}\frac{\sum_{i=M+1}^N{w_i}}{\sum_{i=1}^N{w_i}}\\
=&(x_{rc}^M-x_{rc}^{N-M})\frac{W_{N-M}}{W_N},\\
y_{ct}=&(y_{rc}^M-y_{rc}^{N-M})\frac{W_{N-M}}{W_N}
\end{aligned}
\label{CorVec}
\end{equation}
where $(x_{rc}^{N-M},y_{rc}^{N-M})$ is the WCG of the remaining low-weighted detectors. $W_N$ is the sum of all the weights, and $W_{N-M}$ is the sum of the weights of the low-weighted detectors. Then, we find the reconstructed core location by:
\begin{equation}
\begin{aligned}
x_{rc}&=x_{rc}^M+C_xx_{ct},\\
y_{rc}&=y_{rc}^M+C_yy_{ct}
\end{aligned}
\label{RecCor}
\end{equation}
where $C_x$ and $C_y$ are two constants which have different values in different parts of an array. Optimized values of $M$, $C_x$ and $C_y$ in different regions of an array should be found by examination.
\section{Outside showers identification parameters (OSIP)}\label{sec:OSIP}
Any parameter which can be used as an OSIP should have a different behavior for an OEAS compared with an internal EAS. For example, the mean value function of OSIP should change its trend on a smooth boundary line of an array. But, we should have in our mind that the change of behavior condition (e.g. change of mean value curve trend) of a parameter on the border of the array is a necessary, not an enough, condition. For instance, a parameter which has different mean values in and out of an array may have very wide overlapping distributions inside and outside the array and cannot be used as an OSIP.\\
At first, we begin our consideration with $r_p$ using COG as the core location (as the inventors of $r_p$ have proposed). Our hypothetical array is symmetrical and the density of detectors around every detector of the array is the same in all parts of the array, so we should substitute $\alpha_i=1$ for all $i$s in Eq.~\ref{Rp}.\\
In order to compare the behavior of the mean value function of $r_p$ inside and outside the array, the true core location of each EAS is assumed to be on the line $(i, 0)$ (i increases from 0 (array center, on (0,0)) to $130\textrm{m}$ outside the border line of the array by steps of 1m). In each step, the $r_p$ is averaged for those EASs that satisfy the threshold condition.\\
\begin{figure}
\centering
\resizebox{0.7\hsize}{!}
{\includegraphics[width=\hsize]{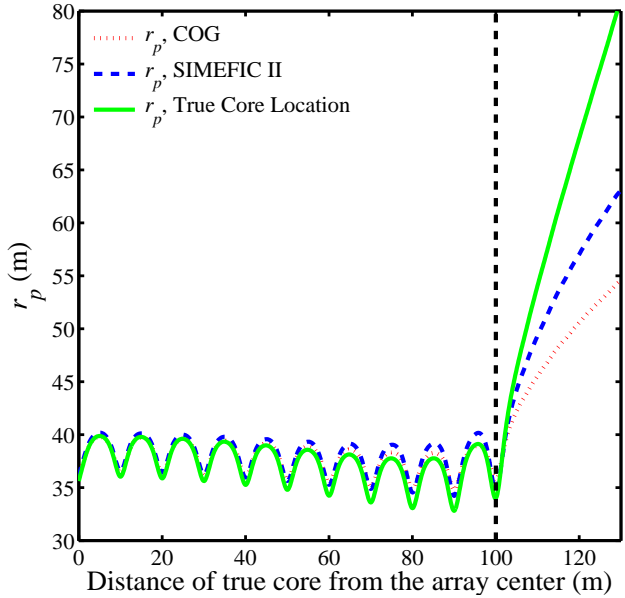}}
\caption{Dotted red line belongs to the average of $r_p$ on horizontal symmetry line of the array for the aforementioned EASs which satisfies threshold condition. The dashed blue line belongs to the results of $r_p$ using SIMEFIC II for the reconstruction of core location. The solid green line belongs to the results of $r_p$ in its ideal form. Border location (side of the array) is depicted by a vertical dashed black line.}
\label{2RpLine}
\end{figure}
As can be seen in Fig.~\ref{2RpLine}, average of $r_p$ (dotted red line) on the border of the array changes its slope and beyond the array border, its mean value quickly increases. So, we expect that a shower with larger $r_p$ is more probable to be an OEAS compared with a shower with smaller $r_p$.\\
In order to optimize the $r_p$ parameter to be used as an OSIP, we reconstruct the core location of the shower by a more precise method than COG. SIMEFIC II method \cite{hedayatiB} has far better results than COG for reconstructing the core location of an EAS. Also, SIMEFIC II method is not so sensitive to the information of an isolated detector with random large pulse height, because it depends on the information of a pair of detectors and it is very impossible for the two detectors far from core location and near each other to have random high-pulse height simultaneously.\\
For reconstructing the core location, SIMEFIC II (M=N/4, Cx=2, Cy=2) which has relatively good precision near the border of the array is used. As can be seen in this figure, when we reconstruct the core location by the SIMEFIC II method (dashed blue line), the slope change of the mean value is even more severe.\\
In Fig.~\ref{2RpLine}, $r_p$ is also shown in its ideal form  (solid green line). In this situation, $r_i$s are distances to the true core location of an EAS (provided by CORSIKA). It shows the ultimate limit for using $r_p$ as an OSIP. As can be seen, SIMEFIC II makes $r_p$ closer to its ultimate behavior in the outside region. Another interesting fact is that, when the core location is outside the array, if someone can reconstruct a core location near the true core location, OEASs can be better identified.
\section{Distribution Functions}
\label{sec:DF}
As said in the previous section, the trend change is only a necessary, not an enough, condition. Another necessary condition for a good OSIP is that its probability distribution should be narrow enough which can efficiently discriminate between a deep OEAS and an internal EAS with the true core location near the boundaries (especially the last internal ring).\\
\begin{figure}[htp!]
\begin{center}
\subfloat[COG is used as the core location.]{%
  \label{3rpCOG}\includegraphics[clip,width=0.5\columnwidth]{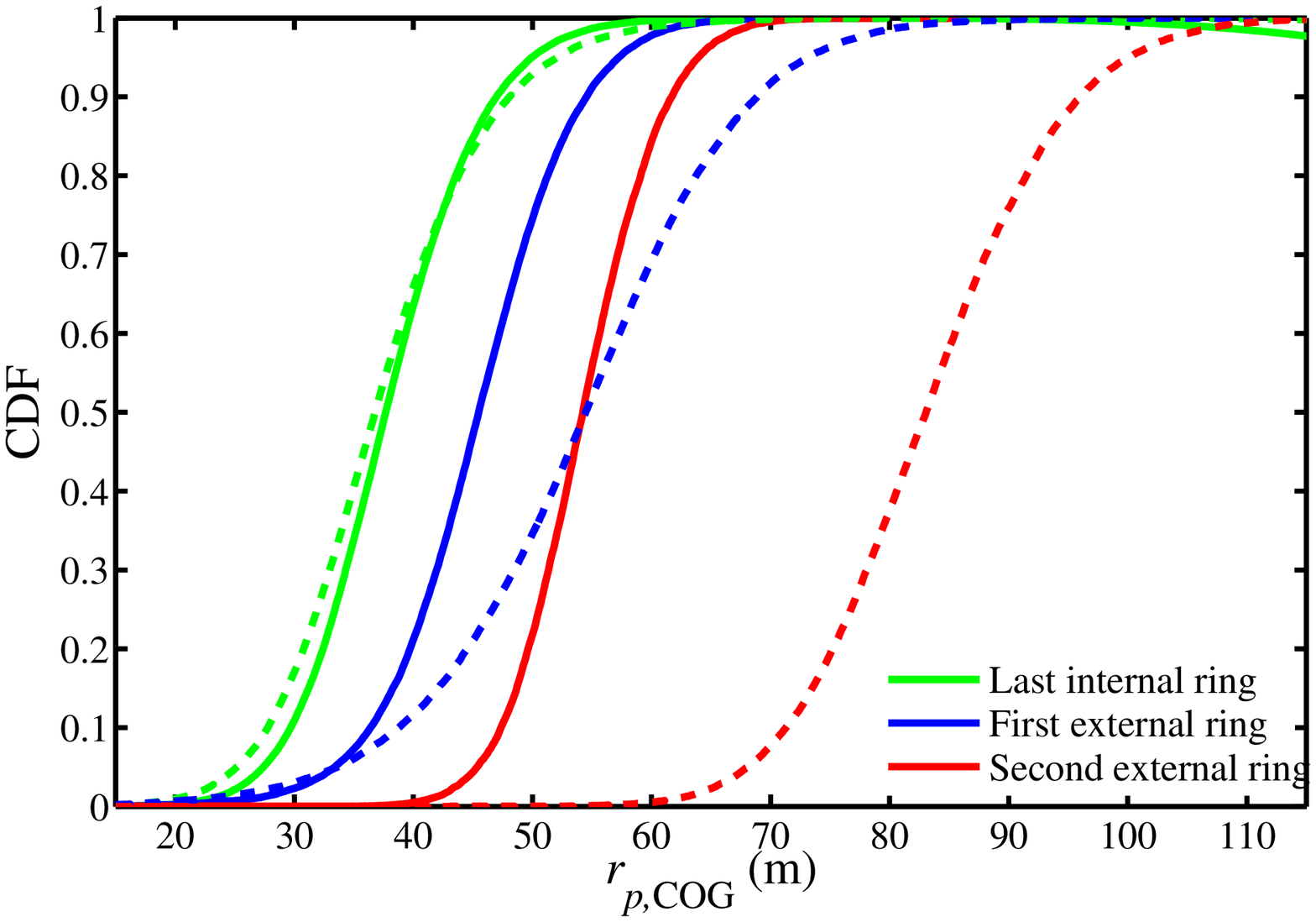}%
}
\subfloat[SIMEFIC II method is used for finding the core location.]{%
  \label{3rpSIMEFIC}\includegraphics[clip,width=0.5\columnwidth]{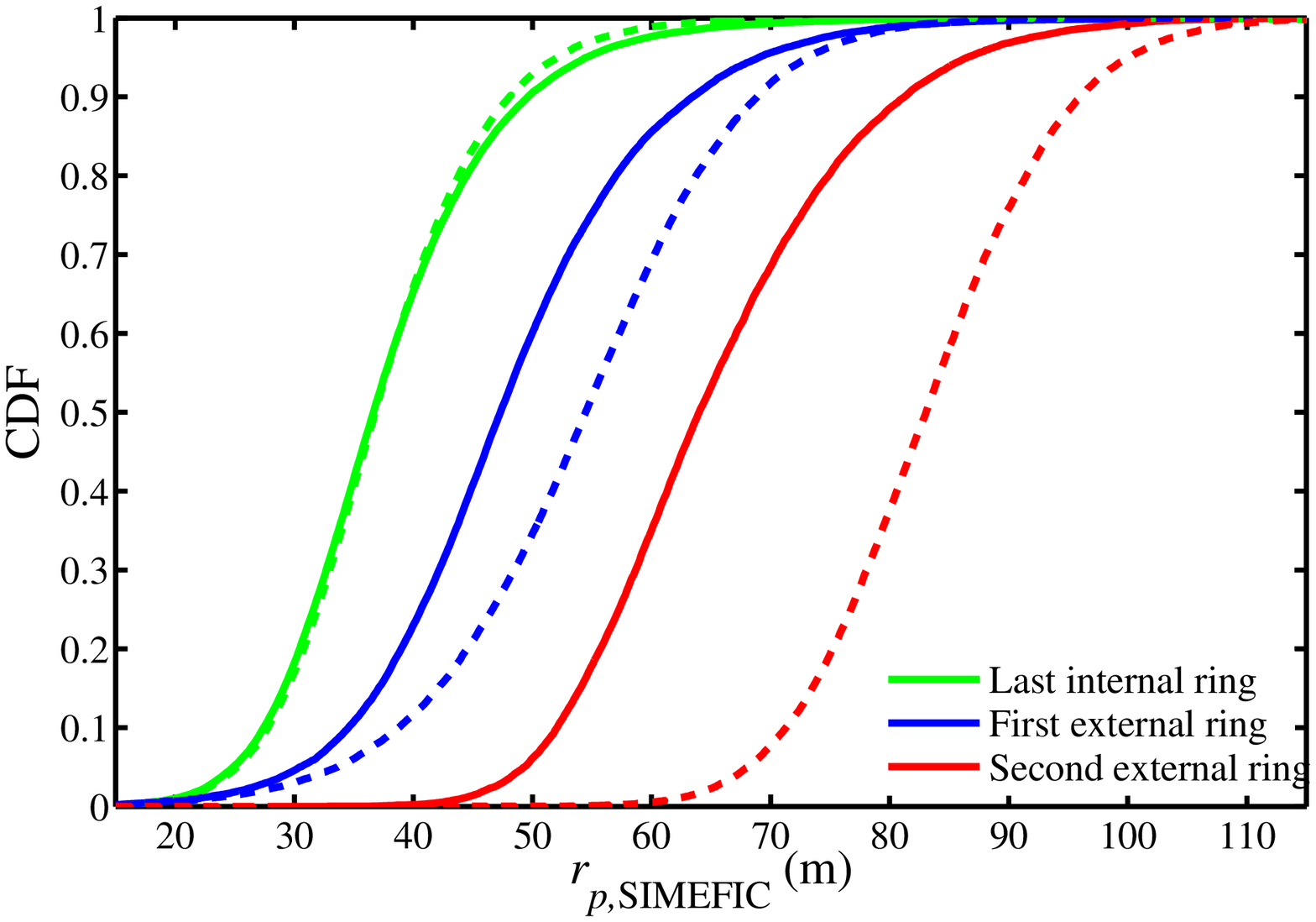}%
}\newline
\caption{CDF of $r_p$ in 3 regions of Fig.~\ref{1_array}. In both figures dashed lines belong to the CDF of the true core location provided by CORSIKA.}\label{RpCDF}
\end{center}
\end{figure}
In order to evaluate the qualification of an OSIP to fulfill this necessary condition in our assumed array, at least an OSIP should discriminate between an EAS with the true core location in the last internal ring of the array (green hatched area in Fig.~\ref{1_array}) and the second external ring (red crosshatched in Fig.~\ref{1_array}). So, we distribute EASs' true core location uniformly in each of the 3 regions shown in Fig.~\ref{1_array} (as mentioned in Sec.~\ref{sec:SimSec}) and, then, evaluate the cumulative distribution functions (CDF) of OSIPs for each area (the true core locations of each EAS are uniformly distributed on each region for 10 times and, whenever an event satisfied the threshold condition, it is applied).\\
Figure \ref{RpCDF}\subref{3rpCOG} shows the results of CDF of $r_p$ using COG as the reconstructed core location, Fig. \ref{RpCDF}\subref{3rpSIMEFIC} shows the results of using SIMEFIC II reconstructed core location for the CDFs of $r_p$. In both parts of Fig. \ref{RpCDF}, dashed lines show the results of $r_p$’s CDFs in its ideal form ($r_i$s are distances from the true core location). It is clear that $r_p$ in its ideal form at most can distinguish the last internal ring's EASs from the second external ring's OEASs nearly completely (when all the second external ring's OEASs are rejected, we can be sure that all OEASs from regions beyond it will be rejected as well (see Fig.~\ref{2RpLine})). But, even when we reject the second external ring completely, we have a large contribution from the first external ring. Also, it can be seen that SIMEFIC II can make $r_p$ a significantly better OSIP than using $r_p$ with COG, because it can hold more events from the last internal ring and, at the same time, reject all the events from the second external ring.\\
\begin{table*}[!t]
\centering
\begin{tabular}{| c | c | c | c |} 
 \hline
\thead{$r_p < r_{p,\textrm{max}}$}& \thead{last internal \\ring} & \thead{first external \\ring} & \thead{second external \\ring}\\ [0.5ex] 
 \hline\hline
$r_{p\textrm{,COG}}< 40$ m & 64\% & 21\% & $\lesssim 0.5\%$ \\
\hline
$r_{p\textrm{,SIMEFIC II}}<42.2$ m & 73\% & 30\% &  $\lesssim 0.5\%$ \\
\hline
$r_{p\textrm{,ideal}}<59.9$ m & 99\% & 69\% &  $\lesssim 0.5\%$ \\
\hline
\end{tabular}
\caption{Some remarkable values extracted from CDF curves in Fig.~\ref{RpCDF}. Column 1 shows the maximum values for each OSIP in order to have less than $0.5\%$ event contamination from the second external ring. Columns 2, 3, and 4 show the percent of accepted events by OSIP from each region.}
\label{table:2}
\end{table*}
Table ~\ref{table:2} shows some interesting values extracted from CDF of the above OSIPs. In this table, we choose the values for the OSIPs whose contamination of OEAS from the second external ring will be less than or approximately equal to 0.5\%. As can be seen, when we optimize $r_p$, the contribution of the last internal ring will be increased by about 9\% compared with the non-optimized $r_p$. At the same time, contamination from the first external ring will increase again by about 9\%.
\section{Core location precision}
Because the area of the first external ring is more than that of the last internal ring, it may seem that those events which will remain from the first external ring in the survived EASs can destroy precision of all the others (especially because of their great error in the reconstruction of core locations). But, because of the correction term in SIMEFIC II method, we can find the core locations of those events from the first external ring with relatively good precision.\\
\begin{figure}
\centering
\resizebox{0.6\hsize}{!}
{\includegraphics[width=\hsize]{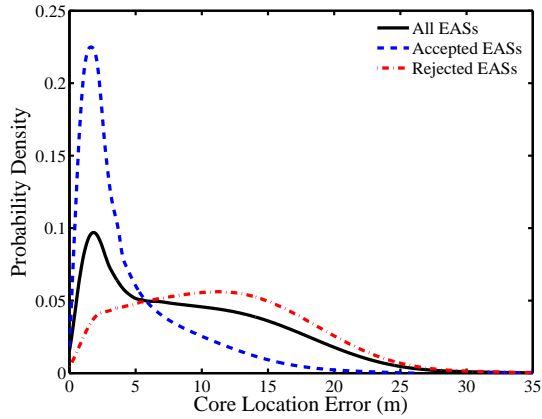}}
\caption{Distribution of the distance between the reconstructed core location and the true axis.}
\label{4core}
\end{figure}
In order to estimate the results of the foregoing technique for the precision of reconstructing the core location, the true core location of the simulated EASs is evenly distributed in all three mentioned regions and, then, the error of core location reconstruction is estimated. The results are shown in Fig.~\ref{4core}. In this figure, distribution of distances of the reconstructed core location by SIMEFIC II ($M=N/4, C_x=2, C_y=2$) from the true axis of EAS (which is a measure of the reconstructed core location error \cite{hedayatiB}) for those EASs which satisfy the threshold condition is shown. As can be seen, when we use the rejection condition ($r_p<42.2\textrm{m}$ using SIMEFIC II for finding $r_i$s), the error occurring in core location reconstruction procedure is far better (with the mean value of 4.4 m). Also, you can see in this figure that the rejected EASs have far worse precision than the selected EASs (the mean value of error is 11.4 m).\\
Actually, with this rejection condition, 32\% of those EASs which are accepted in the threshold condition will remain. The area of the three shown regions altogether is $220 \textrm{m} \times 220 \textrm{m}-190 \textrm{m}\times 190 \textrm{m}=12300\textrm{m}^2$. Therefore, the sensitive area of the array is increased by about $0.32\times12300\textrm{m}^2=3936\textrm{m}^2$. If we compare this area with the area of the array with the border cut of 10 m, the array's area without the last internal ring (that would be $190\textrm{m}\times190\textrm{m}=36100\textrm{m}^2$, an optimistic border cut), the sensitive area of the array will be increased by about 10\%, which is more than the area of the last internal ring ($200\textrm{m} \times 200\textrm{m}-190\textrm{m}\times 190\textrm{m}=3900\textrm{m}^2$). Without this procedure, such an increase in sensitive area needs at least adding a new ring to the array (with an optimistic border cut, rejecting the events with the true core location in this newly added ring), 88 new detectors, and more than $4000 \textrm{m}^2$ increase in the area of the array.\\
If we need higher precision for the reconstructed core location in the perimeter area of the array, we can use a smaller limit value for the $r_p$. However, we should decide that whether we need higher precision or more sensitive area.\\
It may seem that an increase of about 10\% in the sensitive area of the array is not significant, but if we compare the total sensitive area of the array after the application of OSIP cut with the total sensitive area which remains after the application of a border cut in common surface arrays in use around the world, we will discover the usefulness of the OSIP cut. For example, in KASCADE array which was a very sophisticated installation, only EASs with their cores within a radius of 91 m from the center of the array were accepted (i.e. 35\% of the total area of the array was excluded) \cite{antoni2003cosmic}. Therefore, the total effective area added to the array with the method proposed in this paper should not be assumed only 10\%.\\
All of the common reconstruction methods of EASs rely on precise core location estimation. Nevertheless, if we have the core location of marginal EASs with an acceptable precision (as the SIMEFIC II suggests), we can safely reconstruct EASs’ parameters such as energy or arrival direction. For example, to find energy in the Greisen function fitting method, we can use the SIMEFIC II core location estimation as the initial trial core location. Furthermore, as mentioned previously, we should note that marginal EASs are more energetic (the most energetic EASs which can satisfy the threshold condition are those EASs with the true core near the boundary) and, when the core location is known, can be fairly well reconstructed for their higher energy deposit in detectors of the array. For instance, when the initial trial core location is known with a good precision, the estimation of EAS size can be even better for the marginal EASs for their higher energy deposits (see Figures 6 and 8 of \cite{hedayatiA}). Moreover, with the application of a new method introduced for arrival direction reconstruction named SIMAD \cite{hedayatiC}, the arrival direction of marginal EASs can be reconstructed with an acceptable precision.\\
It should be cautioned that above a certain energy threshold, the acceptance of the array is considered to be energy independent. Therefore, above this energy threshold, it is expected that the reconstruction precision of the central EASs is better than the marginal EASs.
\section{Conclusions}
In this paper, an existing technique for the rejection of OEASs was optimized. This technique only relies on the data of array detectors and do not need any other supplemental data such as Cherenkov radiation information of EAS.\\
The most usual method for the rejection of an OEAS is border cut, which significantly decreases the sensitive area of the array and overall efficiency of the array. Also, border cut can decrease the extent of cosmic rays' energy, which can be detected by the array.\\
According to this technique, a parameter called $r_p$ which is the weighted mean distance of triggered detectors from the reconstructed core location is calculated. If $r_p$ for an EAS is more than a certain value, EAS is rejected as an OEAS.\\
Also, it was shown that, if we used a new method for finding core location called SIMEFIC II, we could significantly increase the sensitive area of the array.
In this paper, it was demonstrated that the reconstruction of the core location with better precision could give better results for OEAS rejection procedure. So, the procedure of finding OEASs can be optimized using an optimized version of the SIMEFIC II method for finding core location in an array. Another optimization of the OEAS rejection procedure was possible if one could find another OSIP with a narrower distribution function and or a severe change of mean function trend on the border of the array.
\begin{acknowledgements}
The author wishes to thank the anonymous reviewer for the constructive comments that resulted in a stronger manuscript. 
\end{acknowledgements}


\end{document}